\documentclass[twocolumn,showpacs]{revtex4}
\usepackage{graphicx}
\graphicspath{{pict/}{}}

\usepackage{bm}

\newcounter{Fig}

\newcommand{\be}{\begin{equation}}
\newcommand{\ee}{\end{equation}}

\begin{document}
\title{Fano resonances in magnetic metamaterials}
\author{Uta Naether and Mario I. Molina}
\affiliation{Departmento de F\'{\i}sica, Facultad de Ciencias, Universidad de Chile, Santiago, Chile\\
Center for Optics and Photonics (CEFOP), Casilla 4016, Concepci\'{o}n, Chile}


\begin{abstract}

We study the scattering of magnetoinductive plane waves by internal (external) capacitive (inductive) defects coupled to a one-dimensional split-ring resonator array. We examine a number of simple defect configurations where Fano resonances occur, and study the behavior of the transmission coefficient as a function of the controllable external parameters. We find that for embedded capacitive defects, the addition of a small amount of coupling to second neighbors, is necessary for the occurrence of Fano resonance. For external inductive defects, Fano resonances are commonplace, and they can be tuned by changing the relative orientation/distance between the defect and the SSR array.

\end{abstract}
\maketitle

\section{Introduction}
Metamaterials (MMs) are novel artificial  materials characterized for having negative dielectric permittivity and negative magnetic permeability over a finite frequency range, endowing them with unusual electromagnetic wave propagation properties\cite{MM1, MM2}. One of the most studied MMs is a metallic composite structure consisting of arrays of wires and split-ring resonators (SRRs). The theoretical treatment of such structures relies mainly on the effective-medium approximation where the composite is treated as a homogeneous and isotropic medium, characterized by effective macroscopic parameters. The approach is valid, as long as the wavelength of the electromagnetic field is much larger than the linear dimensions of the MM constituents.
The magnetic properties of SRR arrays have been explored in a number of works \cite{SRR0,SRR1, SRR2,SRR3,SRR4,SRR5,SRR6, SRR7}. 

Among the numerous aspects to explore with a new and promising electromagnetic material such as MMs, are its optical transport properties. For a SRR periodic array, it is only natural to explore the propagation of long-wavelength excitations, in the presence of one or few 
``impurities'' that break its translational invariance. In the absence of any defects,
the eigenmodes of a linear periodic SRR array are termed ``magnetoinductive waves''.
The presence of judiciously placed defects make possible interesting resonance phenomena, 
such as Fano resonances (FR), where there is total reflection of plane waves through the impurity region, in an otherwise periodic potential. In a typical FR system, the wave propagation in the presence of a periodic scattering potential is characterized by open and closed channels. The open channel guides the propagating waves as long as the eigenfrequencies of closed channels do not match the spectrum of linear waves. The total reflection of waves in the open channel occurs when a localized state originating from one of the closed channels resonates with the open channel spectrum\cite{fano}. The FR effect can be used to control the optical response in novel, man-made materials, such as magnetic metamaterials.

In this work, we consider Fano resonance effects due to a few defects whose position, with respect to the periodic SRR array can be easily tuned, making this type of configuration an interesting one to probe experimentally.

This paper is organized as follows: Section II 
reviews the basic equations describing a one-dimensional,
weakly-coupled SRR array in the absence of defects. In section III
we examine a number of simple defect(s) configurations and obtain
the transmission of plane waves across the impurity region, focusing
on possible Fano resonance phenomena and, finally in section IV we
summarize our findings.

\section{Split-ring resonators (SRR) array}

Let us consider a one-dimensional, periodic array of split-ring resonators (SRRs), in the absence of nonlinearity, driving and dissipation. The most simple form of a split-ring resonator is that of a small, conducting ring with a slit. In general, each SRR unit in the array can be mapped to a resistor-inductor-capacitor (RLC) circuit featuring self-inductance $L$, ohmic resistance $R$, and capacitance $C$ built across the slit. In our case, we will consider the case of negligible resistance $R$. Thus, each SRR unit possesses a resonant frequency $\omega_{0}\approx 1/\sqrt{L C}$. In the array, each SRR is coupled to their nearest neighbor via mutual induction\cite{SRR0}.

The evolution equation for the charge $Q_{n}$ residing on the $n$th ring is  
\be
{d^{2}\over{d t^{2}}}\left( L Q_{n} + \sum_{m\neq n} M_{n m} Q_{m} \right) + {Q_{n}\over{C}} = 0\label{eq0}
\ee
where $M_{n m}$ is the mutual induction between rings $n$ and $m$. We cast this equation in dimensionaless form by defining $\tau\equiv \omega_{0} t$, $q_{n}\equiv Q_{n}/C U_{0}$, $\lambda_{n m}\equiv M_{n m}/L$, where $U_{0}$ is a characteristic voltage across the slits.

The dimensionless evolution equation for the charge 
$q_{n}$ residing on the $n$th ring reads now 
\be
{d^{2}\over{d t^{2}}}\left( q_{n} + \sum_{m\neq n} \lambda_{n m} q_{m} \right) + q_{n} = 0\label{eq1}
\ee
where $\lambda_{n m }$ denotes the ratio of the mutual inductances between the $n$th and the 
$m$th ring to the self inductance of the rings, and decreases as the inverse cube of the ring-to-ring distance, $\lambda_{n m} \propto |n - m|^{-3}$ ($m\neq n$). In the limit of weak coupling (large distance between SRR units), it is customary to take $\lambda_{n m}=\lambda\ \delta_{n,m}$.

The two most simple SRR configurations are shown in Fig.1. In one of them (top), all couplings are positive, while for the second (bottom), they are all negative, as a result of Lenz's law. 
\begin{figure}[t]
\includegraphics[scale=0.35]{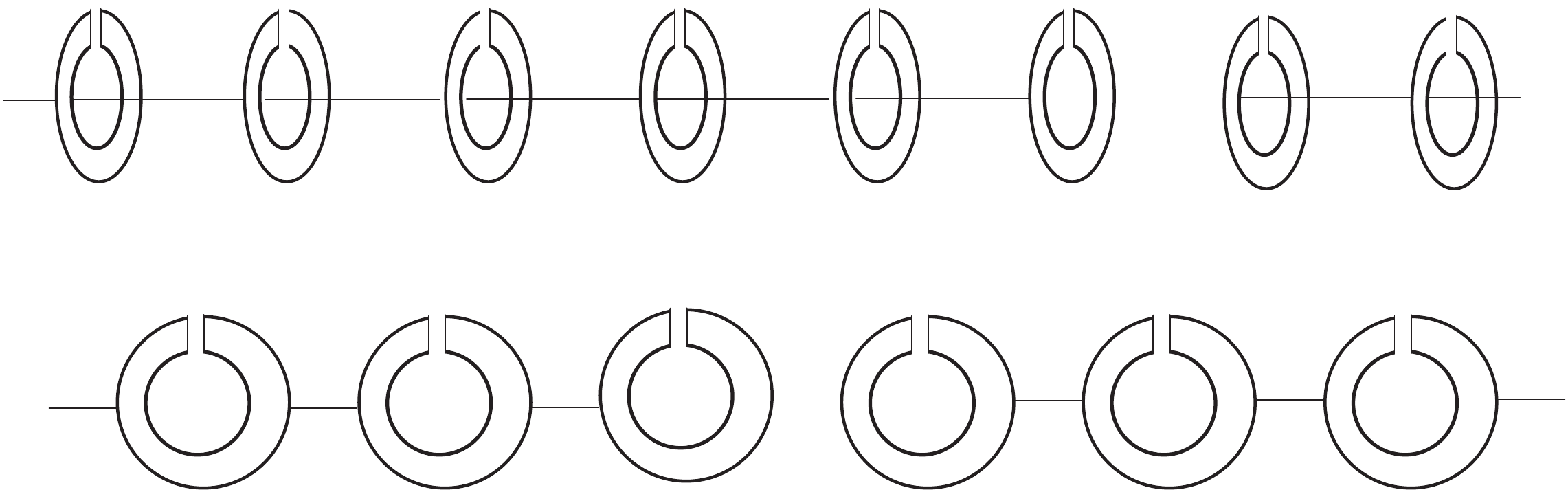}
\caption{Two typical configurations for the SRR array. }
\label{fig1}
\end{figure}
The stationary modes of Eq.(\ref{eq1}) are obtained by posing a solution in the form $q_{n}(t)=q_{n} \exp( i \Omega t)$. This leads to the stationary equations,
\be
-\Omega^{2}( q_{n} + \sum_{m\neq n}\lambda_{n m} q_{m} ) + q_{n} = 0.
\label{eq:3}
\ee 
Magnetoinductive plane waves of the form
$q_{n}=A\ \exp(i k n)$, lead to the dispersion relation
\be
\Omega^{2} = {1\over{1 + 2 \sum_{m>0}\ \lambda_{0 m} \cos(k m)}}.
\ee
We must impose the physical condition $\Omega^{2}>0$. Using that $\lambda_{0 m}$ has the form $\lambda_{0 m} = \lambda/m^{3}$, one obtains:
\be
-{1\over{2 \zeta(3)}} < \lambda < {2\over{3 \zeta(3)}},\label{condition}
\ee
or, $-0.41595 < \lambda < 0.554605$, where $\zeta(s)$ is the Riemann Zeta function $\zeta(s)=\sum_{k=1}^{\infty} k^{-s}$.
As we will see, the limits in condition (\ref{condition}) change a bit when one assumes a weak coupling limit where few or even only one of the $\lambda_{0 m}$ is retained.

\section{Magnetoinductive defects and Fano resonances}

We consider now several simple cases involving defects coupled to the SRR. These defects are created for instance, by altering the electromagnetic characteristics of some  rings, which can  be achieved by changing one or more geometric features of the rings. Another way is by altering their relative coupling with respect to the array. In the case where nonlinearity and dissipation are not considered, two simple choices are to alter the slit width of a ring which alters its capacity only, affecting the value of its resonant frequency. Another simple choice is to couple the array to an external defect(s) consisting of a ring(s) placed outside the SRR array. This affects the value of the coupling between the SRR and the external ring(s) only.

\subsection{Single capacitance impurity}

The first case we study is a single capacitance impurity embedded in the array (Fig.2), which without loss of generality we place at $n=0$. This impurity is created by changing the slit width of the ring at $n=0$. The stationary equation for this case is
\be
-\Omega^{2}( q_{n} + \sum_{m\neq n}\lambda_{n m} q_{m} ) + (1 + \delta_{n,0} \Delta) q_{n} = 0,
\label{single capacitance}
\ee
where $\Delta$ is the change in the (dimensionless) resonance frequency at the defect position. When the slit width tends to zero, its capacitance diverges, making the resonant frequency approach zero. This implies $\Delta\rightarrow -1$. On the other hand, when the slit width is large, the local capacitance goes to zero, and the resonant frequency diverges, implying that $\Delta\rightarrow\infty$. Thus, $-1<\Delta<\infty$.

\begin{figure}[t]
\includegraphics[scale=0.3,angle=180]{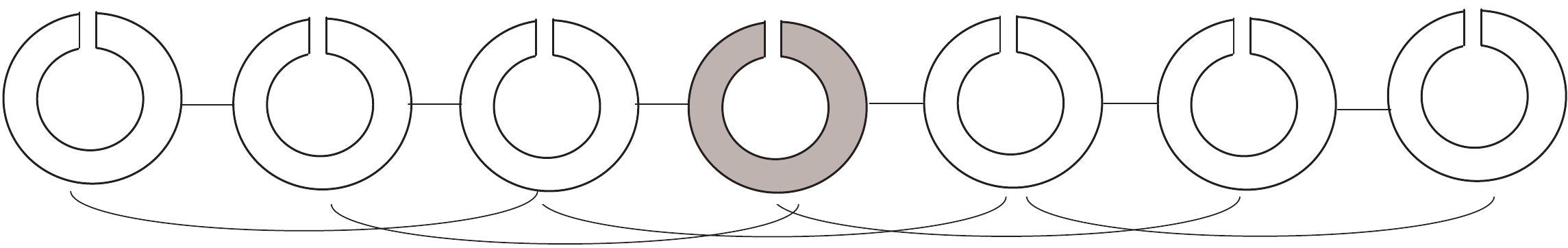}
\\
\includegraphics[scale=0.35]{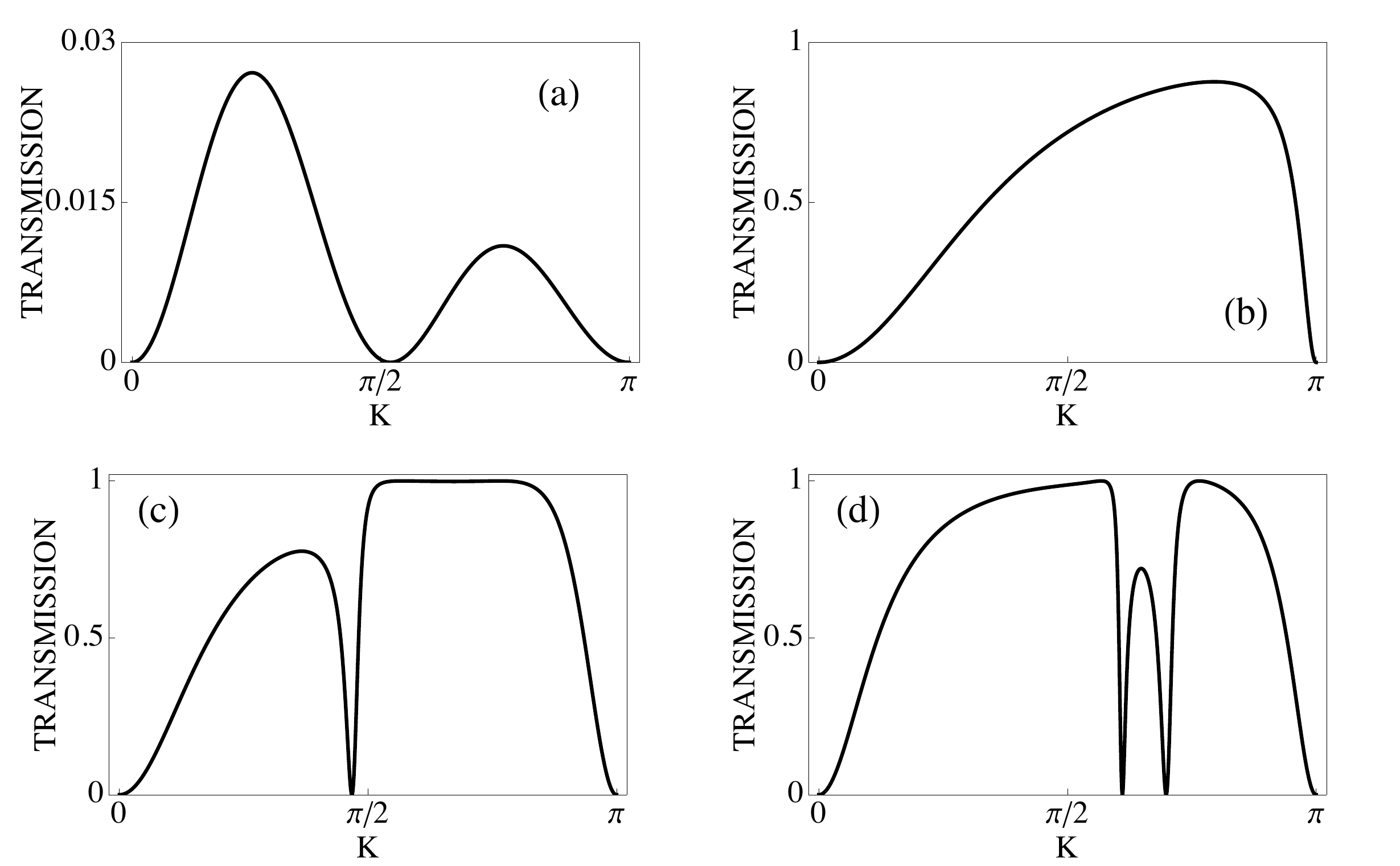}
\caption{Top: Capacitive defect embedded in SSR array, with couplings to first-and second nearest neighbors. Bottom: (a), (b) and (c) show
transmission coefficient vs wavector, for several parameter values: $\lambda=-0.1, \lambda'=(1/8)\lambda,\Delta=0.7$ (a), $\lambda=0.4, \lambda'=0.21, \Delta=-0.5$ (b), $\lambda=0.4, \lambda'=0.21, \Delta=0.5$ (c), $\lambda=0.4, \lambda'=0.21, \Delta=1.0$ (d).}
\label{fig2}
\end{figure}
Usually, embedded defects coupled locally do not lead to Fano Resonance phenomena (perfect plane-wave reflection). In the SRR array however, the couplings are dipolar and therefore, long-range, and non-local effects have to be considered. For computational simplicity we will work with couplings to first-and second nearest neighbors only. The hope is that interference between the path through nearest neighbors and the one through next-nearest neighbors will give rise to Fano Resonance. Equation (\ref{single capacitance}) becomes:
\begin{eqnarray}
-\Omega^{2} \left[ q_{n} + \lambda (q_{n+1}+q_{n-1})+
\lambda' (q_{n+2}+q_{n-2}) \right]&+&\nonumber\\
(1 + \Delta_{n 0}) q_{n} &=&0.\ \ \ \ \label{eq3}
\end{eqnarray} 
where $\lambda'$ is the coupling among next-nearest neighbors. We pose a plane wave solution of the form
\be
q_{n}=\left\{ \begin{array}{ll}
			I\ e^{i k n} + R\ e^{-i k n} & \mbox{$n<-1$}\\
			T\ e^{i k n} &\mbox{$n>1$}
			\end{array}
				\right.
\ee
After replacing this {\em ansatz} into (\ref{eq3}), one obtains after a little algebra:
\be
\Omega^{2} = {1\over{1 + 2\lambda\cos(k) + 2 \lambda' \cos(2 k)}}\label{omega2}
\ee
and
\be
t\equiv |T/I|^2= \left|{A + \Delta\ B\over{A + \Delta\ C}}\right|^2
\label{eq:8}
\ee
where,
\begin{eqnarray}
A&=& -2 i \sin(k)[ 2\lambda'^3\cos(3k)+6\lambda\lambda'(\lambda\cos(k)+\lambda'\cos(2k))\nonumber\\
& &+ \lambda(\lambda^2+3\lambda'^2) ]\nonumber\\
B&=& -2 i \sin(k)[\ 2\lambda\lambda'(\lambda\cos(k)+\lambda'\cos(2k))+\lambda\lambda'\ ]\nonumber\\
C&=&[\ 1+2\lambda \cos(k)+2\lambda'\cos(2k))(\lambda^2+2 \lambda \lambda' e^{-i k}\nonumber\\
& & -\lambda'^{2} + 2 \lambda'^{3} \cos(2k)\ ]
\end{eqnarray}
Now, for the SRR case where dipolar interactions fall with the inverse cube of the distance between SRR units, we have $\lambda'=(1/8) \lambda$. After inserting this into Eqs.(\ref{omega2}) and (\ref{eq:8}) and after imposing $\Omega^{2}>0$, we conclude that $-(4/9) < \lambda < (4/7)$ is the relevant coupling regime for this SRR configuration. Inside this regime it is possible to have a relatively weak Fano resonance, as shown in  Fig.2a. As expected, when coupling to second nearest neighbors is neglected ($\lambda'=0$), there is no FR at all, since in that case the transmission becomes
\be
t = {(2 \lambda \sin(k))^{2}\over{(2\lambda\sin(k))^2 + \Delta^{2}(1+2\lambda\cos(k))^{2}}}, 
\ee
which can only be zero at $k=0,\pi$.
For more general $\lambda,\lambda'$ values, it is possible to have zero, one, and even two different strong Fano resonances, as shown in Fig.2b, 2c and 2d, respectively.
\begin{figure}[t]
\includegraphics[scale=0.3]{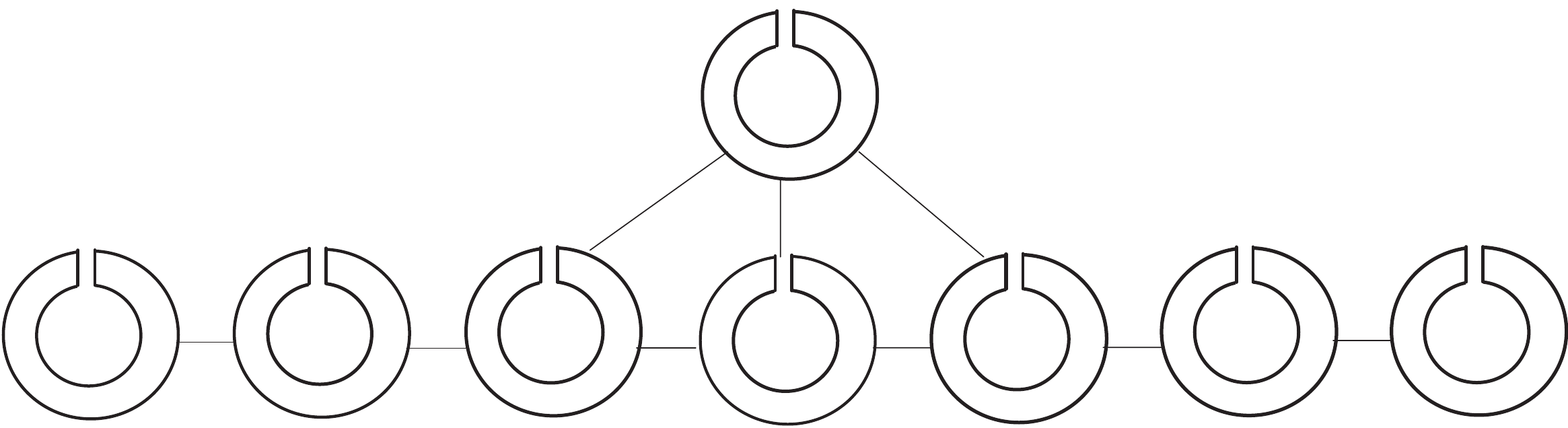}
\includegraphics[scale=0.45, angle=0]{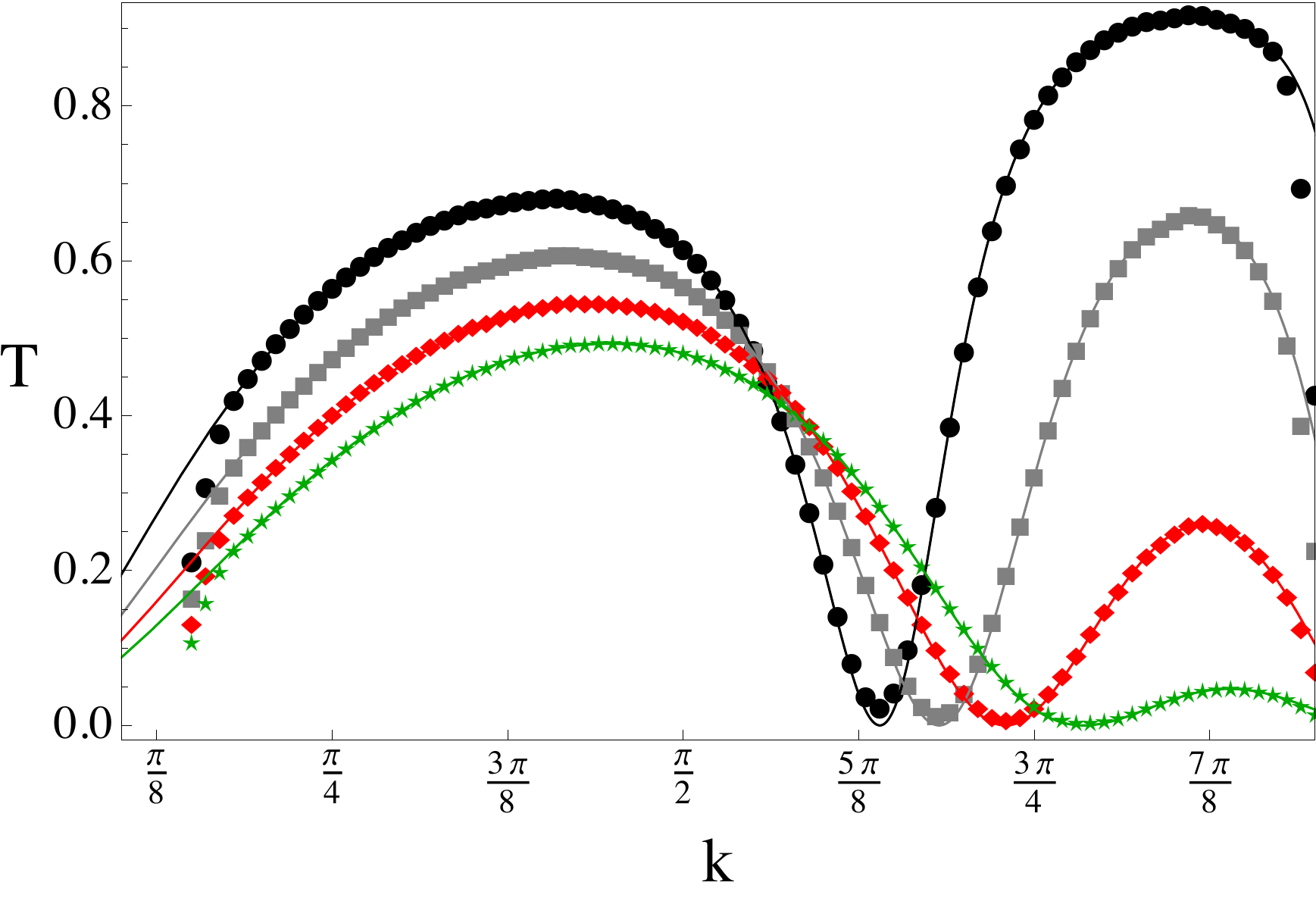}
\caption{(Color online)  Top: SRR array with external inductive defect. Bottom:
Transmission coefficient vs wavevector, for different values of the relative coupling defect: $\lambda_{1}/\lambda=$ 1.25(circles), 1.5(squares), 1.75 (rhombi) and 2.0(stars).}
\label{fig3}
\end{figure}

\subsection{Single inductive impurity}

Another simple case is a single inductive impurity embedded in the array. This is achieved by placing a SRR unit at different distances from the left and right portions of the array, leading to asymmetrical couplings $\lambda_{L}$ and $\lambda_{R}$. Only couplings between nearest-neighbors are asumed. This case was already considered in ref\cite{symms}. In this case, it can be proven (not shown) that there is no FR for any choice of $\lambda_{L}$, $\lambda_{R}$; the transmission vs wavevector curve shows a single maximum at $k=\pi/2$. For the special symmetric case $\lambda_{L}=\lambda_{R}$, there is a transmission resonance ($t=1$) at $k=\pi/2$.

We move now to simple cases where the defect(s) lie(s) outside the SRR array, and thus, their coupling can be tuned by changing their distance to the array. 

\subsection{Single external inductive defect: First case}

The first case of this type is that of a single external inductive defect coupled to the SRR at $n=0$ {\em via} coupling to first and second nearest neighbors via coupling parameters $\lambda_{1}$ and $\lambda_{2}$, respectively (Fig.3). From the geometry of the configuration, it is easy to obtain the relationship
\be
{\lambda_{2}\over{\lambda}}=\left[ 1 + \left( \lambda\over{{\lambda_{1}}}\right)^{2/3} \right]^{-3/2}\label{eqlam2}
\ee
For simplicity, we assume only nearest neighbor coupling among the SRR units. For this approximation to be consistent, one must impose $\lambda_{1}> (1/8)\lambda$ and also $\lambda_{2}>(1/8)\lambda$. From Eq.(\ref{eqlam2}), this implies $\lambda_{1}/\lambda > \mbox{Max}\{(1/8),3^{-3/2}\}=3^{-3/2}=0.192$.

The stationary equations for this case are
\begin{eqnarray}
-\Omega^{2}[ \lambda (q_{n+1}+q_{n-1})&+&q_{n} \nonumber\\
+ q_{e}( \lambda_{1}\delta_{n,0}+\lambda_{2}\delta_{n,\pm 1}) ] + q_{n}&=&0 \nonumber\\
-\Omega^{2} ( \lambda_{1} q_{0}+\lambda_{2}q_{\pm 1}+q_{e} ) + q_{e}&=&0\ \ \ \ 
\end{eqnarray}
where $q_{e}$ is the charge residing on the external defect ring. We assume a plane wave solution of the form
\be
q_{n}=\left\{ \begin{array}{ll}
			I\ e^{i k n} + R\ e^{-i k n} & \mbox{$n\leq 0$}\\
			T\ e^{i k n} &\mbox{$n\ge 1$}
			\end{array}
				\right.
\ee

The transmission coefficient $t\equiv |T/I|^{2}$ for this configuration is
\be
\left|{4 \left( \left({\lambda_{1}\over{\lambda}}\right)\left({\lambda_{2}\over{\lambda}}\right)+\left(1+\left({\lambda_{2}\over{\lambda}}\right)^{2}\right)\cos(k)\right)\sin(k))\over{\left({\lambda_{1}\over{\lambda}}\right)^{2}+4 e^{ik}\left({\lambda_{1}\over{\lambda}}\right)\left({\lambda_{2}\over{\lambda}}\right)+\left(1+e^{2ik}\right)\left({\lambda_{2}\over{\lambda}}\right)^{2}+2i\sin(2k)}}\right|^{2}
\label{Tsingledefect}
\ee
where, $\lambda_{2}$ is given by Eq.(\ref{eqlam2}) for the SRR system.
The most interesting feature of this transmission coefficient is the presence of a Fano resonance, whose position varies with the value of $\lambda_{1}/\lambda$. The FR occurs when $\lambda_{1}\lambda_{2}+(\lambda^2+\lambda_{2}^{2})\cos(k)=0$, which is possible from Eq.(\ref{eqlam2}), for all $0<|\lambda_{1}/\lambda|<2.449$. Therefore, the relevant interval where FR exists in our model is given by $0.192 < \lambda_{1}/\lambda < 2.449$. This can be achieved by simply changing the distance between the SRR array and the external defect. Note from Eq.(\ref{Tsingledefect}) that, if we neglect the effect of $\lambda_{2}$, the position of the FR remains locked at $k=\pi/2$.

To get a better feeling as to what would be expected to see in an actual experiment for this system,
we perform a numerical simulation of a wide pulse impinging on the inductive impurity region. We used an array of $10^{3}$ rings and resort to a symplectic algorithm to trace the time evolution of an initial broad gaussian pulse 
$u_{n}(0)= A \exp(-\alpha(n - n_{0})^{2})\exp(i k (n - n_{0}))$, where $n_{0}=-300$, $A=0.1$ and $\alpha=0.001$ which means a width of about $120$ sites, in order to simulate a plane wave with a well-defined $k$. The rings coupled to the inductive defect are placed at $n=0,\pm 1$. Figure \ref{fig3} shows numerical simulation results for the transmission coefficient vs wavevector for several $\lambda'/\lambda$ values, with $\lambda=0.4$. The agreement between analytics (Eq.(\ref{Tsingledefect})) and numerics is excellent, and the Fano resonances are clearly shown. The numerical discrepancies at wavevectors close to $k=0, \pi$ are due to the long integration times needed since the pulse is quite slow at these $k$ values. In Fig.4 we show the output profile of the gaussian pulse for different $k$-values, after an integration time $t=1500$. The transmission increases monotonically up to $k\approx 0.389\ \pi (0.868\ \pi)$ from the left (right), then decreases steadily until it vanishes completely at $k \approx 0.601\pi$.

\subsection{Single external inductive defect: Second case}

The next case we consider is a variation of the previous case, where the external
defect is now located halfway between two SRR units (Fig.5). Without loss of generality, we take the external defect coupled symmetrically to the units at $n=0$ and $n=1$, with coupling $\lambda'$. As before, the units in the array interact through nearest-neighbor couplings $\lambda$ only. In order for this approximation to be consistent, and taken into account the dipolar nature of the inductive couplings, we need $|\lambda'| > (1/8)\lambda$. 

The stationary equations read
\begin{eqnarray}
-\Omega^{2}(\lambda (q_{n+1}+q_{n-1})+q_{n}+\lambda' q_{e}(\delta_{n,0}+\delta_{n,1}))+q_{n}=0& & \nonumber\\
-\Omega^{2} (\lambda' (q_{0}+q_{1})) + q_{e}=0& &\ \ \ \ 
\end{eqnarray}
\begin{figure}[t]
\includegraphics[scale=0.25, angle=0]{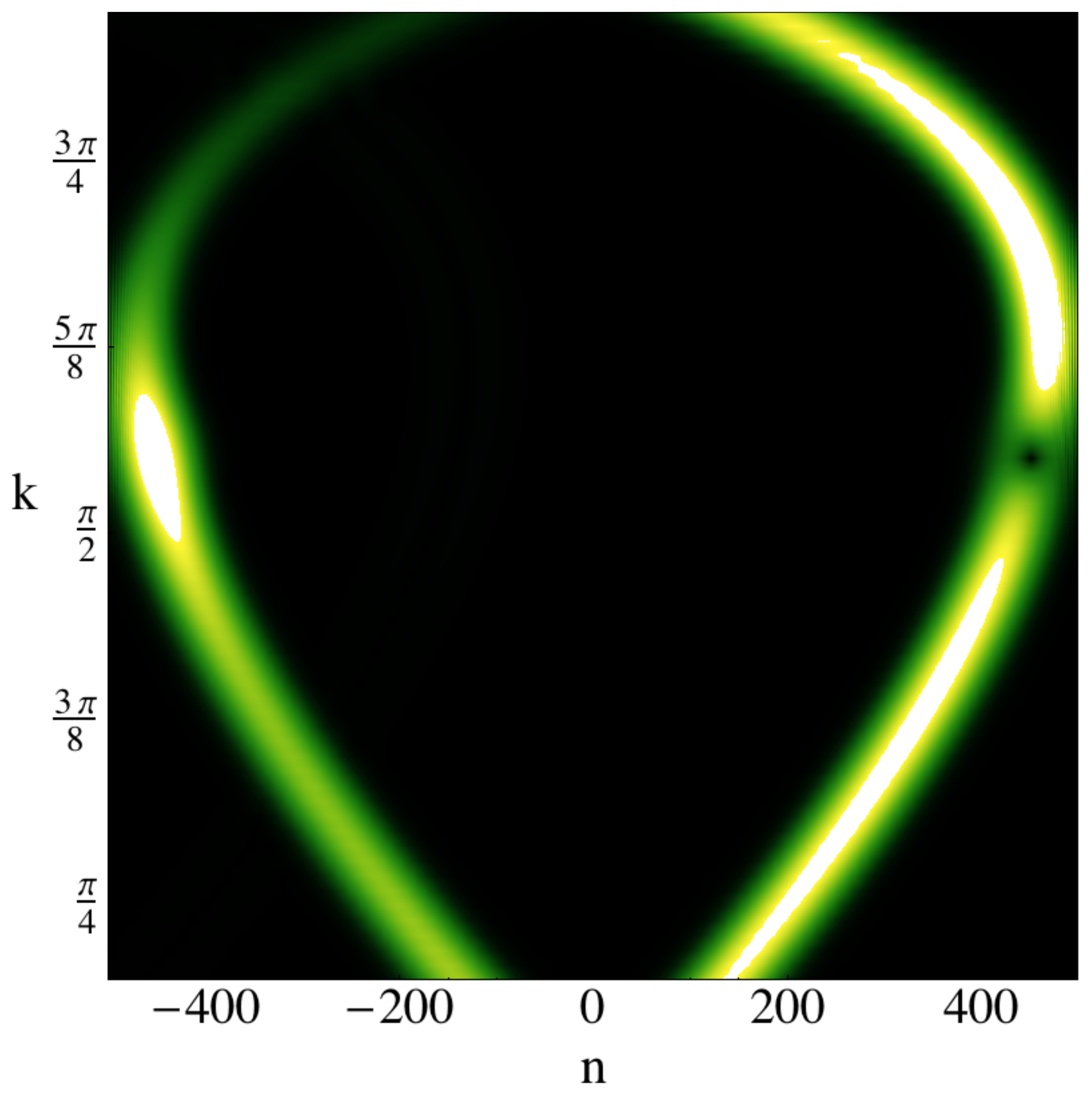}
\caption{(Color online) SRR array with external inductive defect: Output profile after $t=1500$ as a function of wavector and ring site, for $\lambda_{1}/\lambda=1$. Impurity site is located at $n=0$. }
\label{fig4}
\end{figure}
where $q_{e}$ is the charge on the external defect ring. We pose a plane wave solution of the type
\be
q_{n}=\left\{ \begin{array}{ll}
			I\ e^{i k n} + R\ e^{-i k n} & \mbox{$n\leq -1$}\\
			T\ e^{i k n} &\mbox{$n\ge 2$}
			\end{array}
				\right.
\ee
leading to a transmission coefficient:
\be
t \equiv \left|{T\over{I}}\right|^{2}=\left|{(e^{i k} - 1)(\lambda'^{2} + 2 \lambda^{2} \cos(k))\over{2(\lambda'^{2} + \lambda^{2} \cos(k) (1-e^{i k}))}}\right|^{2}.
\ee
Fano resonances are possible when $\lambda'^{2} + 2 \lambda^{2} \cos(k)=0$. This implies, $k_{F}=-(1/2)(\lambda'/\lambda)^{2}$, which is possible only if $|\lambda'/\lambda|<\sqrt{2}$. This constraint plus the consistency condition, give us the relevant coupling parameter window for this SRR configuration:
\be
(1/8) < \left|{\lambda'\over{\lambda}}\right| < \sqrt{2}
\ee
\begin{figure}[t]
\includegraphics[scale=0.3,angle=0]{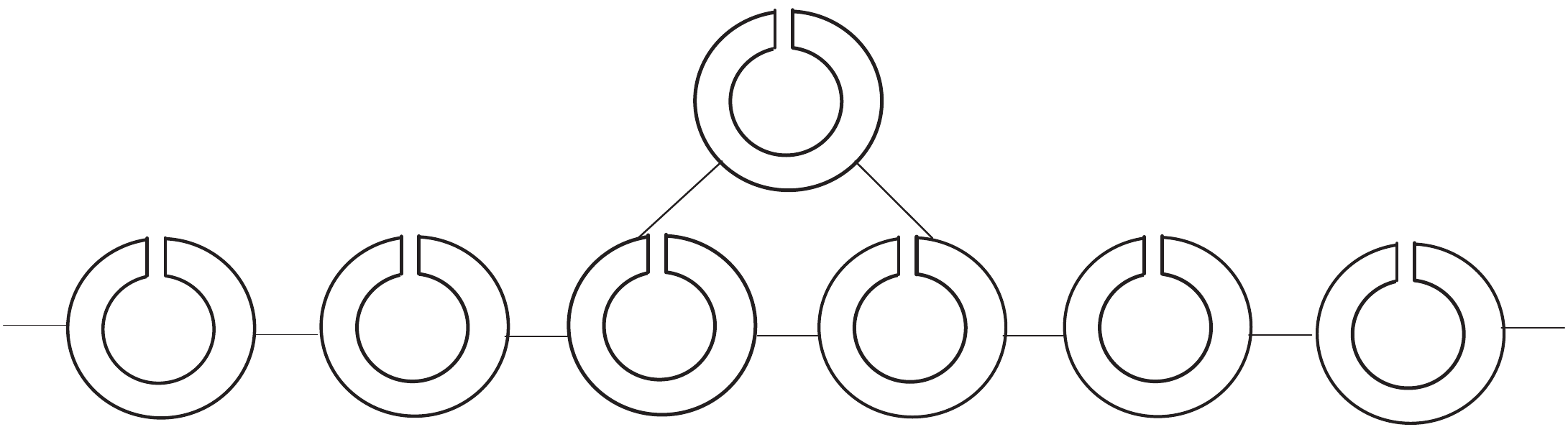}\\
\includegraphics[scale=0.45]{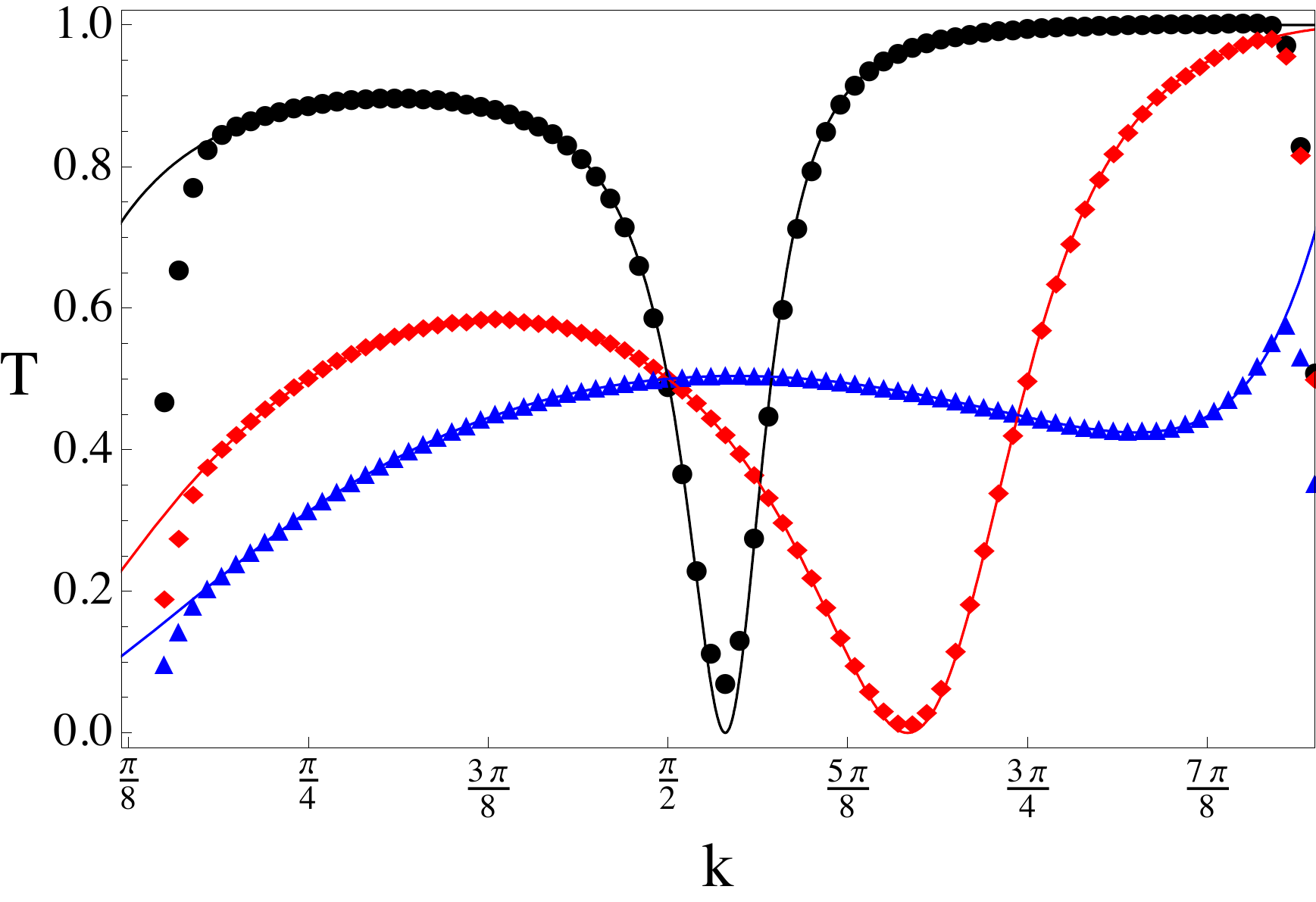}
\caption{(Color online)  Top: External defect coupled symmetrically to the SSR array. Only couplings to first nearest neighbors are considered. Bottom: Transmission vs wavevector for several $\lambda'/\lambda$ values: $0.
5$ (circles), $1.0$ (rhombi) and $1.5$ (triangles).}
\label{fig5}
\end{figure}
\begin{figure}[t]
\includegraphics[scale=0.25, angle=0]{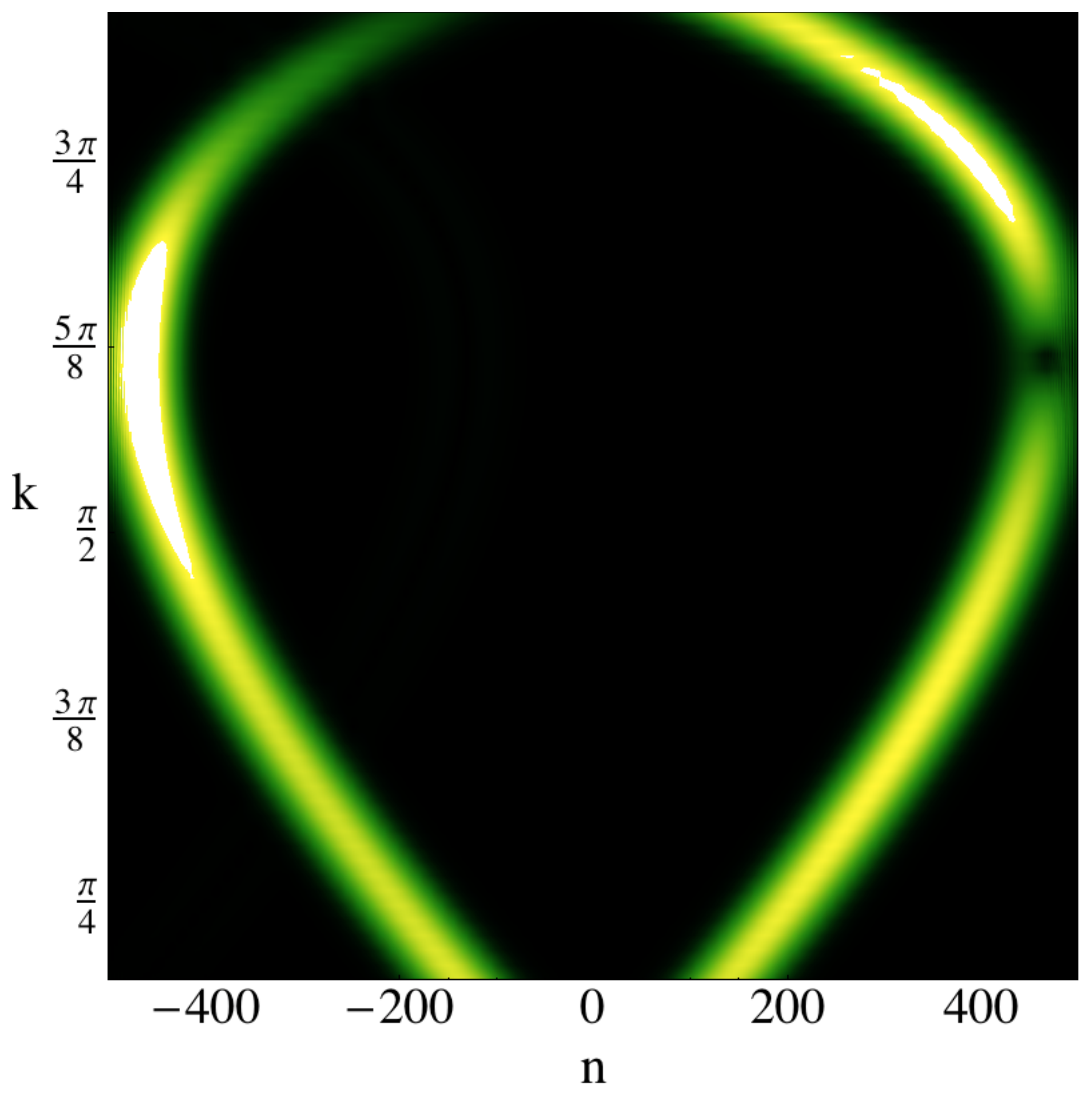}
\caption{(Color online) Fano resonance: External defect coupled symmetrically to
two array units: Output profile after $t=1500$ as a function of wavevector and ring site, for $\lambda_{1}/\lambda=1$.}
\label{fig6}
\end{figure}
Figure 5 shows some transmission curves for parameter values inside and outside this window. The position and width of the Fano resonance depend on the ratio $|\lambda'/\lambda|$, which can be externally tuned by changing the distance between the defect and the array.

Figure 6 shows output intensity profile after propagation of $t=1500$ as a function of wavevector and site position.

\subsection{Two external inductive defects}

\begin{figure}[h]
\includegraphics[scale=0.25,angle=0]{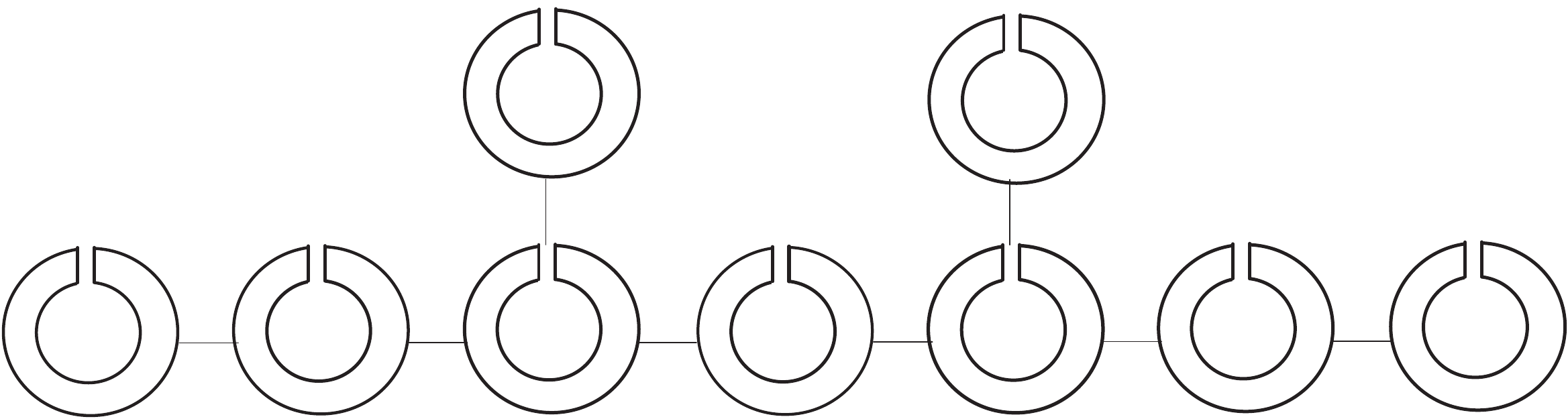}\\
\includegraphics[scale=0.35]{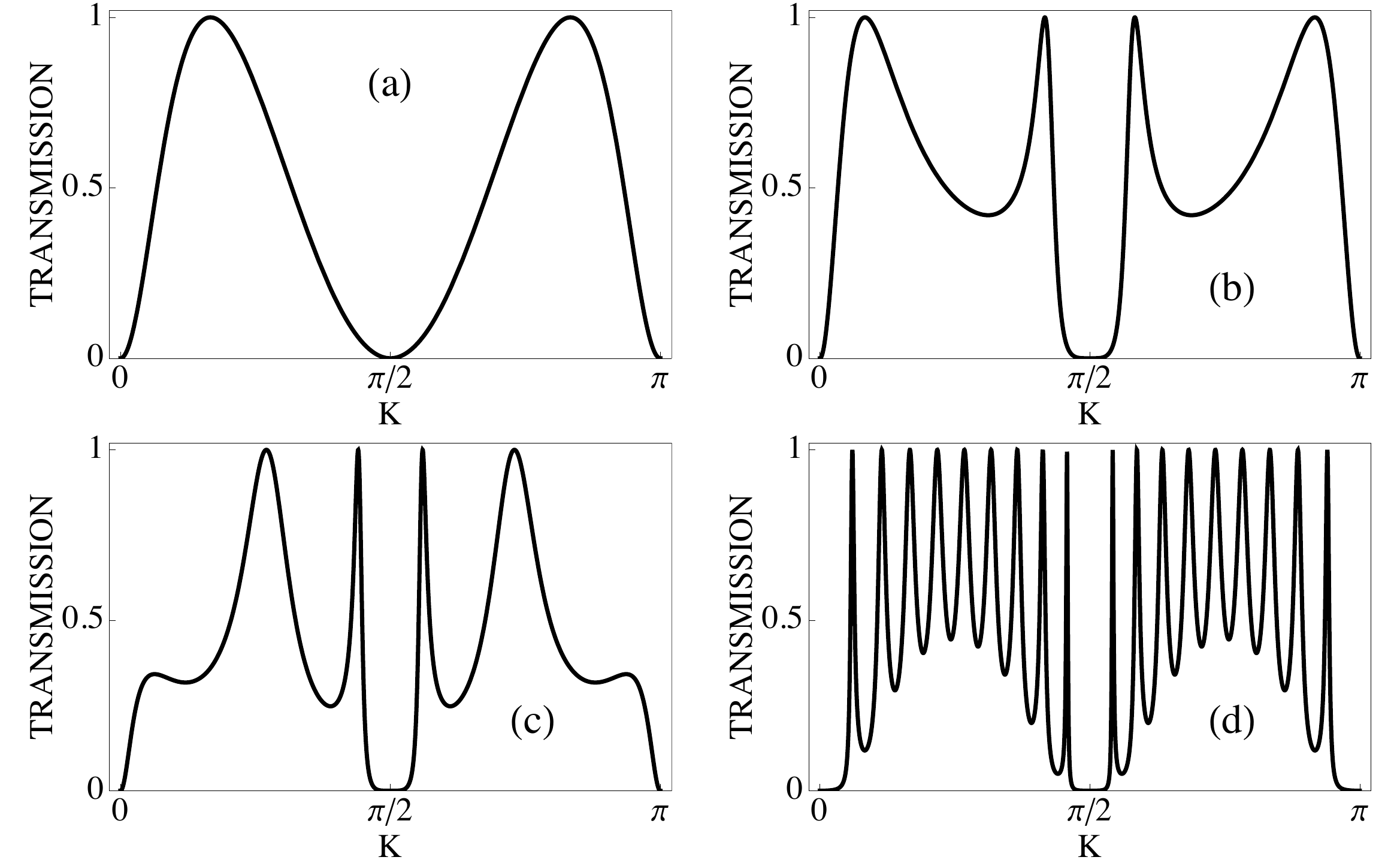}
\caption{Two external defects separated by $L$ units and coupled to the SSR array.  Transmission vs wavevector for several separations $L$: $L=2$(a), $L=3$(b), $L=5$(c) and $L=20$ (d) (in all cases $\lambda=\lambda_{1}=\lambda_{2}=0.4$).}
\label{fig7}
\end{figure}

The final array we consider consists of two external defects, each coupled to a single unit of a SRR array. The units of the array are coupled to first neighbors only via coupling $\lambda$. Between both defects, there are $L$ units (Fig.\ref{fig7}).
Without loss of generality we assume the defects to be coupled to units at $n=0$ and $n=L$, with couplings $\lambda_{0}$ and $\lambda_{L}$, respectively. The stationary equations read
\begin{eqnarray}
-\Omega^{2}[ \lambda (q_{n+1}+q_{n-1})+q_{n}+\lambda_{0} Q_{0}\delta_{n,0}& &\nonumber\\
 + \lambda_{L} Q_{L} \delta_{n,L} ] + q_{n}&=&0  \nonumber\\
-\Omega^{2} (\lambda_{0} q_{0} + Q_{0}) + Q_{0}&=&0 \nonumber\\
-\Omega^{2} (\lambda_{L} q_{L} + Q_{L}) + Q_{L}&=&0 \ \ \ 
\end{eqnarray}
where $Q_{0}$ and $Q_{L}$ are the charges residing at the external rings coupled to units $n=0$ and $n=L$ of the array. As before, we pose a plane wave solution
\be
q_{n}=\left\{ \begin{array}{ll}
			I\ e^{i k n} + R\ e^{-i k n} & \mbox{$n\leq 0$}\\
			T\ e^{i k n} &\mbox{$n\ge L$}
			\end{array}
				\right.
\ee
The transmission coefficient is given in this case by
\be 
\left|{\lambda^{4} (e^{4ik}-1)\over{e^{4ik}[\lambda_{0}^{2}+2i\lambda^{2}\sin(2k)][\lambda_{L}^{2}+2i\lambda^{2}\sin(2k]-\lambda_{0}^{2}\lambda_{2}^{2}e^{2ik (L+2)}}}\right|^{2}
\label{eq:21}
\ee
which is symmetric under the exchange between the external rings, depending only on 
$|\lambda_{0}/\lambda|$, $|\lambda_{L}/\lambda|$ and $L$.
We see from Eq.(\ref{eq:21}) that, regardless of the separation $L$ between the external defects, there is only a single Fano resonance located at $k=\pi/2$.
Figure 7 shows examples of transmission vs wavevector plots for different defects separation. However, it should be noted that, in a more realistic setting that considers the presence of dissipation effects, one would expect to observe some broadening of the resonances, as well as a decrease in their heights, as the separation between defects increases. For SRRs coupled to nearest-neighbors only, this effect is more pronounced at the band edges, and minimized at mid-band\cite{symms}. For couplings beyond nearest-neighbors, the situation is more complex\cite{last}.

\section{Summary}

In conclusion, we have examined Fano resonance effects in a SRR array coupled to internal (capacitive) and external (inductive) defects. In the case of internal or embedded defects, the presence of coupling to second neighbors, in addition to the customary first neighbor coupling, is necessary to induce the FR phenomenon. For external defects, the FR phenomenon depends
mainly of the geometric configuration between the defect(s) and the SRR array. The position and strength of these FR could be easily tuned in current experiments, and constitute a clear example of the possibility of controlling  the transport of electromagnetic waves  across a magnetic metamaterial.

This work was supported by Fondecyt (grant 1080374), Programa de Financiamiento Basal de Conicyt (grant FB0824/2008) and a CONICYT doctoral fellowship.


\begin{thebibliography}{10}

\bibitem{MM1} D. R. Smith, W. J. Padilla, D. C. Wier, S. C. Nemat Nasser, S. Schultz, Phys. Rev. Lett. {\bf 84}, 4184 (2000).

\bibitem{MM2}
J. B Pendry, Phys. Rev. Lett. {\bf 85}, 3966 (2000).

\bibitem{SRR0}
S. Linden, C. Enkrich, G. Dolling, M. W. Klein, J. Zhou, T. Koschny, C. M. Soukoulis, S.  Burger, F. Schmidt, and M. Wegener, IEEE J. Selec. Top. Quant. Electron. 12, 1097 (2006); T. J. Yen, W. J. Padilla, N. Fang, D. C. Vier, D. R. Smith, J. B. Pendry, D. N. Basov, and  X. Zhang, Science 303, 1494 (2004).

\bibitem{SRR1}
M. Gorkunov, M. Lapine, E. Shamonina, K.H. Ringhofer, Eur. 
Phys. J. B {\bf 28}, 263 (2002).

\bibitem{SRR2}
M.C.K. Wiltshire, E. Shamonina, I.R. Young, L. Solymar, Electron. 
Lett. {\bf 39}, 215 (2003).

\bibitem{SRR3}
R.R.A. Syms, E. Shamonina, V. Kalinin, L. Solymar, J. Appl. Phys. 
{\bf 97}, 64909 (2005).

\bibitem{SRR4}
E. Shamonina, V.A. Kalinin, K.H. Ringhofer, L. Solymar, J. Appl. 
Phys. {\bf 92}, 6252 (2002). 

\bibitem{SRR5} 
E. Shamonina, L. Solymar, J. Phys. D-Appl. Phys. {\bf 37}, 362 (2004).

\bibitem{SRR6}
M.J. Freire, R. Marques, F. Medina, M.A.G. Laso, F. Martin, Appl. 
Phys. Lett. {\bf 85}, 4439 (2004).

\bibitem{SRR7}
I. V. Shadrivov, A. N. Reznik and Y. S. Kivshar, Physics B {\bf 394}, 180 (2007).

\bibitem{fano}
For a recent review see, for instance, A. E. Miroschnichenko, S. Flach and Y. S. Kivshar Rev. Mod. Phys. {\bf 82}, 2257 (2010).

\bibitem{symms}
R.R.A. Syms, E. Shamonina and L. Solymar, IEE Proc.-Microw. Antennas Propag., {\bf 153}, 111 (2006).

\bibitem{last}
In the presence of dissipation, a term of the form $i \gamma \Omega$ enters Eq.(\ref{eq:3}). For a complex wavevector $k=k_{r}-i k_{i}$, with $|k_{i}|\ll |k_{r}|$, one obtains $\Omega^{2}\approx (1+\sum_{n m} \lambda_{n m} \cos(m k_{r}))^{-1}$ and $k_{i}\approx \gamma ( 1+\sum_{n m} \lambda_{n m} \cos(m k_{r}))^{1/2}/\sum_{m} \lambda_{n m} \sin(m k_{r})$. Neglect of dissipation effects is valid provided $|k_{i}| L\ll 1$.

\end{thebibliography}
\end{document}